\documentclass[12pt]{article}
\usepackage{amssymb,amsmath,esint,titlesec,mathrsfs}
\titleformat*{\section}{\normalsize\bf}
\titleformat*{\subsection}{\small\bf}

\begin{document}


\begin{titlepage}

\setlength{\baselineskip}{18pt}

                               \vspace*{0mm}

                             \begin{center}

{\Large\bf  Logarithmic and power-law entropies from convexity}

                                   \vspace{45mm}

              \normalsize\sf  NIKOLAOS \  \  KALOGEROPOULOS $^\dag$\\

                            \vspace{1mm}

                   {\small\sf  Center for Research and Applications \\
                                  of Nonlinear Systems  \  \  (CRANS),\\
                          University of Patras, Patras 26500, Greece.\\ }
                                
                                      \end{center}

                            \vspace{35mm}

                     \centerline{\large\bf Abstract}
                                
                                     \vspace{3mm}
                     
    \noindent In an attempt to understand the origin and robustness of the Boltzmann/Gibbs/Shannon 
  entropic functional, we adopt  a geometric approach and discuss the implications of the 
  Johnson-Lindenstrauss lemma and of Dvoretzky's theorem on convex bodies for the choice 
  of this functional form. We contrast these results with a more recent result on flowers of balls,
  which may be interpreted as suggesting the use of power-law entropies for some systems.\\

                                             \vfill


                             \vspace{0mm}

\noindent\rule{12.5cm}{0.3mm}\\  
   \noindent   {\small\rm $^\dag$        Electronic mail: \ \ \   {\sf nikos.physikos@gmail.com}} \\
   
\end{titlepage}
 

                                                                                \newpage                 

\rm\normalsize
\setlength{\baselineskip}{18pt}

\section{Introduction} 

The Boltzmann/Gibbs/Shannon (BGS) functional has been very successful in describing the collective behavior of systems in thermodynamic equilibrium.
What is more imrpessive is that it is also used successfully, almost exclusively, in predicting the behavior of systems out of equlibrium, or even in equilibrium 
but not resembling the ideal gas which provided a motivation for its formulation, such as systems with long-range interactions. There are numerous other 
functionals which have been proposed over the years, such as the Renyi entropy, as alternatives to the BGS functional, many motivated by considerations in information theory. \\

During the last three 
decades some attention has been paid to the q-entropies as well to a multitude of other entropic functionals, many of them having the same general power-law form as the q-entropies,
which are motivated from and oriented toward Statistical Mechanics.  At this point of development, it is probably fair to say that none of these functionals can match the breadth, the scope 
and the success of the BGS functional.  Actually, one could claim that  most these more recently proposoed  functionals do not appear to have any relation to Physics, as they have not 
been able to explain or unambiguously predict the behavior of any  physical systems which have been studied in an experimentally verifiable way.  
Moreover, most of these functionals lack the theoretical/dynamical foundations that the BGS entropy enjoys, and their use relies on conjectures or data fittings,
none of which are solid ground on which to build a robust statistical mechanical framework.  These drawbacks are certainly not enough to 
invalidate these functionals, but rather they present the challenge of investigating their foundations, which would inevitably lead us to understand better the the BGS functional itself. 
To avoid a case-by-case analysis, we focus on the general distinction of functional forms: the BGS entropy, representing functiomal forms involving the logarithm, vis-a-vis the q-entropies as 
representing power-law functionals. We consider the examination of more complicated functionals, such as combinations of the above forms, outside the scope of the present work, 
for reasons of simplicity. \\    

To try to understand this distinction, we rely on geometric considerations, borrowed from Convex Geometry, and more 
specifcally from Asymptotic Geometric Analysis. In Section 2 we present the Johnson-Lindenstrauss lemma and its implications, according to our viewpoint, 
for entropy. In Section 3 we look at entropy from the viewpoint of Dvoretzky's theorem for convex bodies and for flowers, suggesting a possible origin of 
power-law entropies. In Section 4 we present conclusions and an outlook for further work.\\ 
       

\section{The Johnson-Lindenstrauss lemma and entropy}

The general reference for several aspects of Asymptotic Geometric Analysis which we need is the two-volume \cite{AGM1}, \cite{AGM2} 
where one can find all background material, proofs and much more, which we omit in order to keep the exposition focused and reasonably short.  \\

The Johnson-Lindenstrauss ``flattening'' lemma is a statement in Convex Geometry having numerous applications, especially in Computer Science and data processing.
Consider \ $\mathbb{R}^d$ \ endowed with its Euclidean norm \ $\| \cdot \|$ \ and a set of \ $n\in\mathbb{N}$ \ points in it, indicated by \ $x_1, x_2, \ldots , x_n \in \mathbb{R}^d$. \ 
The goal is to see whether this configuration of points can ``fit'' inside a linear subspace \ $\mathcal{X}$ \ of \ $\mathbb{R}^d$ \ of lower dimension \ $k\in\mathbb{N}$ \ without distorting 
their distances ``too much''. We notice initially that if \ $d\geq n$ \ then this is trivially possible as all \ $x_1, \ldots, x_n$ \ can be chosen to be in \ $\mathcal{X}$ \ by using as
embedding \ $\phi: \mathbb{R}^d \rightarrow \mathcal{X}$ \ the identity, which is an isometry. So the question is non-trivial when \ $d<n$.  \ 
The Johnson-Lindenstrauss ``flattening'' lemma \cite{JL} states  that such a linear embedding \ $\phi$ \ exists and the dimension \ $k$ \ varies as \ $\log n$.  \    
With the above notation, more precisely, let \ $0<\epsilon <1$ \ and 
\begin{equation}
    k \geq \frac{c}{\epsilon^2} \log n
\end{equation}
Then there is a linear map \ $\phi: \mathbb{R}^d \rightarrow \mathbb{R}^k$ \  such that for all \ $i\neq j, \ \ i,j = 1,2,\ldots, n$ \ 
\begin{equation} 
(1-\epsilon) \| x_i-x_j \| \leq \| \phi (x_i) - \phi (x_j) \| \leq (1+\epsilon) \| x_i-x_j \| 
\end{equation}
The map \ $\phi$ \ can be  chosen to be multiple of an orthogonal projection onto \ $\mathbb{R}^k$ \ as in the original proof  of the lemma \cite{JL}.
The distances are allowed to be distorted at most by an absolute amount \ $\epsilon$, \ according to (2).  
What is important for this work is the logarithmic dependence of \ $k$ \ on the number of points \ $n$. \        
It should be stressed that \ $k$ \ is independent of the dimension of \ $\mathbb{R}^d$. \  It only depends on the number \ $n$ \ of points considered, and on the 
``magnitude'' \ $\epsilon$ \ of the  allowed distance distortion.  The map \ $\phi$ \ is Lipschitz, actually bi-Lipschitz, which implies that \ $\phi$ \  
is almost everywhere differentiable, according to Rademacher's theorem. This is desirable in most cases in Physics where  
maps are assumed to be infinitely differentiable, unless there is the possibility to prove, or a reason to demand, lower regularity. \\  

The pertinent question is what is the relation between the Johnson-Lindenstrauss lemma and the BGS entropy. Our view and proposal is that the BGS entropy has the logarithmic form
which allows it to be robust and widely applicable exactly because of the validity of the dimensional reduction provided by the Johnson-Lindenstrauss lemma. 
To be concrete, let us assume that we are dealing with a conservative system whose underlying dynamics is Hamiltonian which eventually determines its thermodynamic behavior. 
To bridge the gap from microscopic dynamics to thermodynamics one invokes the concept of ``coarse-graining'' \cite{Coarse}. 
This is clear in both views of L. Boltzmann and J. W. Gibbs, despite their quite distinct approaches to entropy. The question on how to justify and perform such a coarse-graining is 
quite imporant as it determines the resulting probability distribution used in the statistical analysis of the system under study. \\

At this point, we  have to acknowledge the views of 
Landau and Khintchine which consider the knowledge of such microscopic details as irrelevant in determining the thermodynamic behavior of a system, and instead rely on the large 
number of degrees of freedom to deduce the probability distribution determining the statistical behavior of the system \cite{Coarse}. A criticism against this viewpoint is that limit theorems  
which are used in determing the probability distributions usually rely on assumptions, such as having independent random variables or having a distribution with  finite moments etc, 
which may not necessarily be valid or be unnecessarily restrictive for the systems under study.  Another criticism is that one may not necessarily wish to study a system with a huge number
of degrees of freedom, or have a system where taking the thermodynamic limit may not obvious or even desireable \cite{Gross}. This is not to necessarily invalidate the Landau-Khintchine viewpoint, 
but suggest another view on this matter, which may conjecturaly apply to systems not described by the BGS entropy, 
and which considers as important the underlying dynamics of the system in determining its thermodynamic behavior. \\

The equations of fundamental Physics and many equations that are used for effective models rely on Hamiltonians which are quadratic in their canonical momenta. Therefore, we will 
focus from now on to such systems, except where is explicitly stated otherwise.  The ``potential'' part of the Hamiltonian 
usually starts having a quadratic term in the degrees of freedom and proceeds adding interaction terms which are usually of some low-degree polynomial form and which are treated as 
perturbations of the quadratic/``free'' theory. In essence, in such models, one deals with perturbations of harmonic oscillators. Harmonic oscillators 
of a given energy are represented by ellipsoidal-shaped ``unit spheres'' in phase space.  One may start therefore by understanding the behavior of these ``unit spheres''
and subsequently introduce the role of interactions as ``distortions'' of these spheres, at least initially. Following the conventional approach, we assume a large number of degrees of freedom
as the current arguments are better applicable to such cases. \\    

The phase space of a finite-dimensional Hamiltonian system is not, in general, a linear space, but has the structure of a symplectic manifold. Symplectic manifolds do not possess local invariants
except their dimension, unlike Riemannian ones, due to Darboux's theorem \cite{McDS}. This may be a reason behind the unexpected effectiveness of linear techniques in symplectic geometry
as indicated in \cite{Ostrover}, for instance. At any rate, since the tangent space of a symplectic (or any smooth)  manifold is a linear space, understanding of the phase space dynamics of a system is related to understanding some of its linear algebraic behavior. As a result, if we want to understand the Hamiltonian evolution of a system, we may want to start by understanding the 
behavior of the ``unit sphere'' of the system in the ``linear'' approximation in phase space.\\                     

So,  let us consider a system of \ $d/2 \in\mathbb{N}$ \ degrees of freedom having as phase space \ $\mathbb{R}^d$.  \ 
Take \ $n=d$ \ linearly independent vectors of \ $\mathbb{R}^d$ \ on the unit sphere of \ $\mathbb{R}^d$. \ Interactions will deform this sphere, as stated above.
So, the Euclidean distances between these points will change as the system evolves but probably not too abruptly, or not too much, for sufficiently small times.
 The entropy functional we are seeking should ``average out''  details of the evolution in \ $\mathbb{R}^d$, \ in order to provide a description in terms of few thermodynamic parameters.
 From a metric viewpoint these details, which are not very important in the thermodynamic description of the system,  can be seen as distance distortions which do not exceed a 
 particular bound \ $\epsilon$. \  We want to see whether  the configurations of these \  $d$ \ vectors can be reduced to a simpler, lower-dimensional, configuration within the acceptable distance distortion.            
 According to the Johnson-Lindenstrauss ``flattening'' lemma \cite{JL} this is indeed possible.  The resulting configuration belongs to a linear subspace of dimension of  order \ $\log d.$ \               
Hence, if the coarse-graining is seen as the operation which ignores distortions within \ $\epsilon$ \, then the entropy will achieve the desired reduction in the required details of the system as 
long as it is expressed as a logarithm of the original number of vectors \ $d$, \ which is exactly  what the functional form of the BGS entropy does. \\   

A question that arises  is whether such a reduction is possible for other norms on \ $\mathbb{R}^d$, \ or whether it is a unique feature of the Euclidean norm.
From a physical viewpoint this question is related to effective Hamiltonians, usually phenomenologically introduced, which involve kinetic terms which are not quadratic in the canonical momenta,
or kinetic equations, such as the Fokker-Planck equation with fractional derivatives or having non-linear terms etc \cite{Fr}. In all cases examined the results have been negative, indicating that 
the Johnson-Lindenstrauss reduction is applicable to, nearly-Hilbert spaces and not to general Banach spaces, such as \ $L^p$ \ spaces, with \ $\ p\in [1, 2)\cup (2, \infty]$ \   \cite{LMN}. \        
The answer to the question was given in \cite{JN}. Let \ $c_2(\mathcal{Y})$ \ denote the  Euclidean distortion of a finite dimensional normed space \ $\mathcal{Y}$. \ 
This is defined to be the infimum of all constants \ $C>0$ \ such that  there is a linear map \ $\psi: \mathcal{Y} \rightarrow l^2 $ \  satisfying 
\begin{equation}
 \| x \| \leq \| \psi(x) \| \leq C \| x \|,  \ \ \ \ \forall \  x\in\mathcal{Y} 
\end{equation}
One can immediately see that \ $c_2(\mathcal{Y})$ \ is the Banach-Mazur distance \cite{AGM1} between \ $\mathcal{Y}$ \ and the corresponding Hilbert space \ $l^2$. \  
 Moreover, for \  $z\in [1, \infty)$ \ let  \ $\log^\ast (z)$ \  be the unique \ $s\in \mathbb{N}$ \ such that if \ $a_1 =1$ \ and iteratively \ $a_{j+1} = e^{a_j}$ \  then 
 \ $a_s  < z \leq a_{s+1}$. \ With this notation, \cite{JN} prove the following: Let \ $\mathcal{B}$ \ be a Banach space such that for every \ $n\in\mathbb{N}$ \ 
 and for every \ $x_1, \ldots, x_n \in \mathcal{B}$ \ there exists a linear subspace \ $\mathcal{V}$ \  of dimension at most \ $A \log n$, \ where \ $A, B>0$ \ are constants,             
 and a linear map \ $\varphi:   \mathcal{B} \rightarrow \mathcal{V}$  \ such that         
 \begin{equation}
      \| x_i - x_j \| \leq \| \varphi (x_i) - \varphi (x_j) \|    \leq  B\| x_i - x_j \|, \ \ \ \ \forall \  i,j \ \in \{1,\ldots, n \}
 \end{equation}
 Then for every \ $m\in\mathbb{N}$ \ and for every \ $m$-dimensional subspace \ $\mathcal{X} \subset \mathcal{B}$ \ 
 \begin{equation}            
        c_2 (\mathcal{X}) \leq 2^{2^{c \log^\ast (m)}}
 \end{equation}
 where \  $c(A,B) >0$ \ is a constant. 
  This theorem proves that a normed space which satisfies the Johnson-Lindenstrauss lemma  is very close to being a Hilbert space in that all its $m$-dimensional subspaces are isomorphic to Hilbert 
 space with maximum distortion given by (5). \\
 
 In the original proof of the Johnson-Lindenstrauss lemma \cite{JL}, the linear map \ $\varphi : \mathcal{B} \rightarrow \mathcal{V}$ \ was taken to be the multiple of an orthogonal projection. 
 Projections have a special significance in Statistical Mechanics as they involve an ``integration along each fiber'' \ $\mathcal{W}_y = \varphi^{-1}(y), \ \ y\in \mathcal{V}$ \ which replaces the 
 degrees of freedom of the ``fiber'' \ $\mathcal{W}_y$ \ with their averaged contribution to the quantities of interest. In essence they are the geometric picture of the marginal distributions and are seen in numerous occasions, such as the BBGKY hierarchy which relies exactly on a series of such projections.  
 To be clear, here \ $\varphi^{-1}$ \ denotes the inverse image, rather than the inverse map.   \\       
 
 It is understood that even though nearly-Hilbert spaces satisfy the Johnson-Lindenstrauss lemma, it is also possible that other Banach spaces which are not Hilbert spaces 
  satisfy the lemma too as long as they violate condition (5). A pertinent theorem was proved in \cite{JN} to which we refer for further details and proofs of the above statements.\\
 
 It may also worth exploring whether the Johnson-Lindenstrauss lemma is valid for non-linear maps. This question can be quite relevant for our purposes,
 if we go back to considering Hamiltonian evolution in general symplectic manifolds rather than confine ourselves to the case of linear spaces as we have done in this Section. 
It seems however  that only negative results for non-linear maps are known, see \cite{JN} and references therein, so we will  not pursue this matter any further but restrict ourselves to the linear
algebraic  case.\\   
 
 It seems that the Johnson-Lindenstrauss ``flattening'' lemma provides a justification about the success of the logarithmic form of the BGS entropy. This is probably true, if one interprets  the BGS  entropy
 as a functional which is ``averaging out'' rendering the details of bounded distance distortions of the unit sphere in the phase space of the system as irrelevant for the thermodynamic behavior of the system.
 This view of entropy appears to be quite different from the Kolmogorov-Sinai definition of metric entropy \cite{KH}, which is used extensively in dynamical systems.  But one might want to notice that in the case of the Komogorov-Sinai entropy the logarithmic form is put by hand, whereas in what we propose it appears naturally/``dynamically'' 
 as part of the general framework of the dimensional reduction. 
 If one constrains himself to short times in order to remain within the linear space approximation, and chooses a partition of the phase space \ $\mathbb{R}^d$ \ by near translates of the 
 convex ``unit sphere'', and uses a logarithmic form of entropy, 
 then it is possible that these two definitions of entropy might give comparable results. Ultimately this is unclear and warrants further investigation. \\
 
 The reduction in dimension of the Johnson-Lindenstrauss  lemma  is only known for linear spaces and convex bodies.
 It is unknown what would be applicable if the effect of interactions cannot be
 expressed as small perturbations, so the resulting ``unit spheres'' may become non-convex.  At this point we notice there are different degrees of failure of convexity. 
 A star-shaped object, for instance,  is a relatively  mild  departure from convexity.  An extensive theory of star bodies exists paralleling the theory convex bodies \cite{Gardner}. 
 Things become more manageable if additional properties are required of star bodies.  Then it may be possible the ``reduction in complexity'' which is expressed through the logarithmic reduction in
 dimension in the Johnson-Lindenstrauss lemma may turn out to have a non-logarithmic form for such star bodies, as is pointed out in the next Section. 
 This could constitute the beginning of an argument about the origin and domain of validity of power-law entropies in some Hamiltonian systems.\\
 

\vspace{5mm}

\section{Dvoretzky's theorem and entropies; from spheres to flowers}

It may be worth changing our perspective at this point. Since there is no obvious extension of the Johnson-Lindenstrauss lemma to Banach spaces that are not nearly Euclidean,
we may have to widen the framework a bit and look under a different angle into the possibility of a conjectured reduction starting from some general normed space to a Euclidean one. \\

From a physical viewpoint, most theories have a metric structure. Most of the times the corresponding distance function   comes from a norm, as in the cases of Riemannian, Finslerian etc manifolds.
We usually work with mostly Riemannian manifolds, maybe allowing for some form of singularities. 
On the other hand, given the abundance of models used for various purposes, using non-Euclidean norms is a possibility that should not be dismissed off hand. \\
  
  Dvoretzky's theorem is a central result in convex geometry \cite{AGM1}, \cite{AGM2}. One could argue that V. Milman's proof  of Dvoretzky's theorem 
  through the concentration of measure phenomenon on the sphere, has been even more influential in Convex Geometry. 
  We mention in passing, that from a physical viewpoint, the concentration of measure states a fact which makes the connection of Statistical Mechanics with equilibrium
  thermodynamics possible: that functions which depend on many variables almost none of which vary ``too fast'' are essentially constant  \cite{AGM1}.  
  Going back to Dvoretzky's theorem,  let \ $\epsilon >0$ \ and \ $k\in\mathbb{N}$. \ 
  Then there exists an \ $N = N(k,\epsilon )$ \ such that whenever \ $\mathcal{V}$ \ is a normed space  of dimension \ $n\geq N$, \ one can find a  $k$-dimensional subspace
  \ $\mathcal{X}$ \ of \ $\mathcal{V}$ \ such that   
  \begin{equation}
   d_{BM}(\mathcal{X}, l^2_k) \leq 1+\epsilon
  \end{equation}
The distance function \ $d_{BM}$ \ in Dvoretzky's theorem refers to the Banach-Mazur distance, and \ $l^2_k$ \ denotes the \ $k$-dimensional linear space endowed 
with the Euclidean norm\\

In essence, Dvorezky's theorem states that a high dimensiomal normed space, has a subspace with nearly Euclidean sections. The conclusion may be interpreted as a 
mathematical justification of the ubiquidity of the Euclidean norms in physical models, at least at the mesoscopic and macroscopic levels. 
For, even if  one starts with a model endowed with a non-Euclidean distance function related to a norm, 
then upon reduction to a mesoscopic or macroscopic scale, the result will be close to a Euclidean or Riemannian model. 
All this is true within the linear approximation, as we are considering only linear spaces and not general symplectic manifolds as phase spaces.\\

In more geometric terms, the theorem states states that every high dimensional centrally symmetric convex body, which is the unit sphere in a normed space, 
has central sections which are almost ellipsoidal.  The greatest interest for our purposes lies in the dependence of \ $k$ \ on \ $n$. \ 
The answer is provided by the  following Dvoretzky-Milman 
theorem \cite{AGM1}: Let \ $\mathcal{V}$ \  be an $n$-dimensional normed space. Then for every \ $0<\epsilon <1$ \  there is a \ $k\in\mathbb{N}$, \ given by 
\begin{equation}
     k \geq c\epsilon^2 \log n
\end{equation}
with \ $c>0$ \ constant and a $k$-dimensional subspace \ $\mathcal{X} \subset \mathcal{V}$ \ such that \ $d_{BM} (\mathcal{X}, l^2_k) \leq 1+\epsilon$ \ 
where \ $d_{BM}$ \ indicates the Banach-Mazur distance. 
It is exactly this logarithmic dependence of \ $k$ \ on \ $n$ \ which, in our opinion, makes the use of a functional involving a logarithm such as the BGS entropy
so sucessful in describing hte thermodynamic behavior of many systems in equilibrium.  \\

It should be noted that Dvoretzky's theorem and the 
subsequent logarithmic dependence of \ $k$ \ on \ $n$ \ rely crucially on staying within the class of convex bodies throughout.  
However, as was pointed out above, this is just a linear space approximation to the full Hamiltonian evolution which is of interest, when the effect of the interactions is considered 
as negligible. One would like to perform a similiar analysis with the interactions present. Not totally unexpectedly, this is not practically feasible for several reasons. 
One reason is that when interactions are involved, one does not have any strong condition such as convexity which allows us to reach any results. \\ 

The next best set of objects that can be sufficiently analyzed is that of star bodies as mentioned at the end of last Section. 
These are objects which have ``some convexity''. Even though they are not sufficiently general to describe Hamiltonian evolutions, they are probably rigid enough to reach some desireable results. It turns out though, that for our purposes even this class of objects is unmanageable.  We need  to consider 
star bodies having additional properties which will allow us to draw some pertinent conclusions.  
One such property is the convexity of the distance function. In Riemannian manifolds, this is a property typical of complete non-compact manifolds 
which have non-positive sectional curvature, as complete compact manifolds do not admit any non-trivial convex functions. 
Given the abundance of metrics of non-positive curvature for Riemannian manifolds, or for
more general metric spaces in a comparison sense \cite{Gr}, it may not be too much to impose the same convexity condition for 
the radial function function of the star bodies in Euclidean space which we try to analyze.
Such objects have indeed been defined, are currently under study, and are called flowers \cite{MMR}, \cite{MR}. \\

One definition of flowers is that they are the star bodies whose radial function is convex \cite{MMR}.  
It turns out that an equivalent, and more synthetic definition is the following: let \ $B(x, r)$ \ denote a ball centered at \ $x\in \mathbb{R}^n$ \ of radius \ $r$. \  
Let \ $\mathcal{B}_0$ \ indicate the  set of all balls containing the origin, and let \ $B_x = B(\frac{x}{2}, \frac{|x|}{2})$ \ be balls having \ $[0,x]$ \ 
as their diameter.  A flower is any set of the form \ 
\begin{equation}
\mathcal{F} = \bigcup_{i\in I} B_i
\end{equation}
of balls \ $\{B_i \}_i \subseteq \mathcal{B}_0$. \ 
It turns out that there is an analogue ot Dvoretzky's theorem for flowers, which goes as follows \cite{MR}. 
Let \ $\mathcal{F} \subset \mathbb{R}^n$ \  be an origin-symmetric  flower. Then for every \ $\epsilon>0$ \ there is   
a subspace \ $\mathcal{E} \subset \mathcal{F}$ \  whose dimension is \ $c(\epsilon) n$ \ such that   
\begin{equation}
    d(P_\mathcal{E} \mathcal{F}, B^2_n \cap \mathcal{E}) \leq 1+\epsilon 
\end{equation}
where \ $P_\mathcal{E}$ \ indicates orthogonal projection onto \ $\mathcal{E}$,\  and \ $B^2_n= B(0,1) \subset \mathbb{R}^n$. \ 
It may be worth mentioning at this point that if one replaces projections with sections in Dvoretzky's theorem for 
flowers, the theorem is no longer true \cite{MR}, unlike the better known case of convex bodies. \\

If one accepts the interpretation of entropy as a functional which ``averages out'' features of a model so that its effective description is reduced to a model 
possessing far fewer degrees of freedom, and which is endowed with a Euclidean metric, then we see that for flowers this reduction is not very dramatic, 
as in the case of convex bodies where it is logarithmic, but instead it is linear in the dimension of the original phase space. 
Since flowers are not convex bodies, if we assume that they may be close, in some models, to describing ``unit spheres'' after taking into account interactions, 
then it is possible that the use of power-law, and not just linear, entropic functionals such as the q-entropies 
may be required to describe such systems  instead of employing the BGS entropic functional, which would be unsuitable in these cases. \\   

Flowers have another useful property: they are the complements of spherical inversions of convex bodies containing the origin \cite{MMR}, \cite{MR}. 
Apart from their geometric significance, flowers accommodate  duality properties which may provide a better understanding of the dynamical foundations of entropy.
Indeed, one can go to the quantum level and observe that the duality expressed through the polar inversion of convex bodies, more specifically ellipsoids, may 
``pave the way for a  geometric and topological version of quantum indeterminacy'' \cite{Gosson1}. \ Given the special role of Gaussians as 
minimum uncertainty wavefunctions, and the close dynamical  relation of coherent states with the classical harmonic oscillator, the latter of which was used 
throughout the present work through its ``unit sphere'', further exploring this polar inversion, whose natural setting is that of flowers, may be of some interest
for understanding better the behavior of the collective behavior of systems described by logarithmic or power-law functionals.  \\   
     

\vspace{5mm}

\section{Conclusions and Discussion}

In this work we used the Johnson-Lindenstrauss ``flattening'' lemma and Dvoretzky's theorem for convex bodies and flowers in an attempt to understand the origin and reasons for 
the success of the logarithmic functional form of the BGS entropy, and explore possible dynamical features of systems described by power-law entropies. This work follows the spirit 
of the treatment of our earlier  \cite{NK}.
We worked  initially within the linear space approximation which gave rise to convex ``unit spheres''   and only at the very end we incorporated the lack of convexity which should be due
to interactions in a very particular context, by using the recent construction of flowers. Our approach has common points with different approaches to ``coarse-graining''. We consider
the closest one in spirit, if not in details, the approach through ``quantum blobs'' \cite{Gosson2} which may be referring to quantum Physics but it has very strong connections to linear 
symplectic maps and  semi-classical quantization through minimal ellipsoids in classical phase spaces.\\    

In the present work we  interpreted the entropy as a controlled distance deformation, rather than 
as a quantity involving optimal covers of the phase space compatible with a Hamiltonian evolution. The relation between these two interpretations as well as with the 
ideas of L. Boltzmann or J.W. Gibbs about entropy, all of which share the same functional form, is unclear.\\

Our arguments are hand waving, and even though we rely on solid mathematical results, our interpretation allows for many possible objections and questions to be raised. 
Apart from the obvious, which is the validity of this intepretation itself, the biggest issue which emerges is that of the appearence and possible significance of dualities of the convex or 
star bodies  appearing as ``coarse-graining'' cells in classical Statistical Mechanics \cite{Coarse} or as minimum uncertainty cells in quantum systems \cite{Gosson1}, \cite{Gosson2}, \cite{Gosson3}. Such dualities may be behind the possible ``covariant'' behavior of the q-entropies under M\"{o}bius transformations of their q-parameter, as indicated in \cite{GT}. \
One could use such constructions to further probe aspects of the Hamiltonian evolution in symplectic manifolds, and in particular determine the role, if any, 
of the symplectic non-squeezing theorem and the associated symplectic capacities \cite{McDS} in the statistical description of a system of many degrees of freedom.
The current abundance, and ever increasing number of duality constructions in Convex Geometry \cite{ASW}, and the close relation between Sympletic and Convex 
Geometry \cite{Ostrover}, may provide a set of tools which may help resolve some of these issues.   \\        


\vspace{10mm}

\noindent{\bf Acknowledgement:} \ We would like to thank Professor P. Benetatos for numerous discussions on Statistical Mechanics over many years,  
for providing encouragement and support, as well as several references, and Professor E. Vagenas for providing some pertinent references. \\


                                                                                                              \vfill


\newpage

\begin{thebibliography}{99}
\bibitem{AGM1} S. Artstein-Avidan, A. Giannopoulos, V.D. Milman, \ \emph{Asymptotic Geometric Analysis, \  Part I}, \ \ Math. Surveys Monog. {\bf 202}, \ Amer. Math. Soc., \ \ 
                               Providence, RI, USA  \ (2015).
\bibitem{AGM2} S. Artstein-Avidan, A. Giannopoulos, V.D. Milman, \ \emph{Asymptotic Geometric Analysis, \  Part II}, \  \ Math. Surveys Monog. {\bf 261}, \ Amer. Math. Soc., \ \ 
                              Providence, RI, USA  \  (2021).
\bibitem{JL} W.B. Johnson, J. Lindenstrauss, \ \emph{Extensions of Lipschitz mappings into a Hilbert space}, \ pp. 189-206  in \emph{Conference in Modern Analysis and Probability, 
                                New Haven, Conn. 1982}, \ Comtemp. Math. {\bf 26}, \ Amer. Math. Soc., \ Providence, RI, USA \ (1984).                                                           
\bibitem{Coarse} P. Castiglione, M. Falcioni, A. Lesne, A. Vulpiani, \ \emph{Chaos and Coarse Graining in Statistical Mechanics}, \ Cambridge Univ. Press, \ Cambridge, UK \ (2008).
\bibitem{Gross}  D.H.E. Gross, \ \emph{Microscopic statistical basis of classical thermodynamics of finite systems}, \ {\sf arXiv:cond-mat/0505242 [cond-mat.stat-mech]}
\bibitem{McDS} D. McDuff, D. Salamon, \ \emph{Introduction to Symplectic Topology, \ Third Edition}, \ Oxford Univ. Press, \ Oxford, UK \ (2017). 
\bibitem{Ostrover} Y. Ostrover, \ \emph{When symplectic topology meets Banach space geometry}, pp.  959-982, in \emph{Proceedings of the International Congress of Mathematicians, Seoul 2014, 
                                   Volume II},  \ S.Y. Young, Y.R. Kim, D.-W. Lee, I. Yie (Eds.),  \ Kyung Moon Sa Co. \ Seoul, Korea \ (2014).
\bibitem{Fr} T.D. Frank, \ \emph{NonLinear Fokker-Planck Equations: Fundamentals and Applications}, \ Springer Series in Synergetics, \ Springer-Verlag, \ Berlin, Germany \ (2005). 
\bibitem{LMN}  J.R. Lee, M. Mendel, A. Naor, \ \emph{Metric structures in $L_1$: dimension, snowflakes and average distortion}, \ Eur. J. Combin. {\bf 26}(8), \ 1180-1190 \ (2005). 
\bibitem{JN} W.B. Johnson, A. Naor, \ \emph{The Johnson-Lindenstrauss Lemma Almost Characterizes Hilbert Space, But Not Quite}, \ Discrete Comput. Geom. {\bf 43}, \ 542-553 \ (2010).
\bibitem{KH} A. Katok, B. Hasselblatt, \  \emph{Introduction to the Modern Theory of Dynamical Systems}, \ Cambridge Univ. Press, \ Cambridge, UK \ (1995).
\bibitem{Gardner} R.J. Gardner, \ \emph{Geometric Tomography, \ Second Edition} \ Cambridge Univ. Press, \ Cambridge, UK \ (2006). 
\bibitem{Gr} M. Gromov, \ \emph{Sign and geometric meaning of curvature}, \ Rend. Semin. Mat. Phys. Milano {\bf 61}, \ 9-123 \ (1991).
\bibitem{MMR} E. Milman, V. Milman, L. Rotem, \ \emph{Reciprocals and Flowers in Convexity}, \ pp. 199-227 in \ \emph{Geometric Aspects of Functional Analysis; Israel Seminar (GAFA) 
                             2017-2019, Volume II}, \ B. Klartag, E. Milman, (Eds.), \ Springer Nature, \ Switzerland \ (2020). 
\bibitem{MR}  V. Milman, L. Rotem, \ \emph{Novel view on classical convexity theory}, \  J. Math. Phys. Anal. Geom. {\bf 16}(3), \ 291-311 \ (2020).        
\bibitem{Gosson1} M. De Gosson, \ \emph{Quantum polar duality and the symplectic camel: a new geometric approach to quantization}, \ Found. Phys. {51}(3), \ 1-39 \ (2021). 
\bibitem{NK} N. Kalogeropoulos, \ \  \emph{Entropies from coarse-graining: convex polytopes vs. ellipsoids}, \\ 
                       Entropy {\bf 17}, \ 6329-6378 \ (2015). 
\bibitem{Gosson2} M.A. de Gosson, \ \emph{Quantum Blobs}, \ Found. Phys. {\bf  43}, \ 440-457 \  (2013).                    
\bibitem{Gosson3} M. De Gosson, F. Luef, \ \emph{Symplectic capacities and the geometry of uncertainty: the irruption of symplectic topology in  classical and quantum mechanics}, \
                                  Phys. Reports {\bf 484}(5), \ 131-179 \ (2009).   
\bibitem{GT} J.P. Gazeau, C. Tsallis, \ \emph{M\"{o}bius Transforms, Cycles and q-triplets in Statistical Mechanics}, \ Entropy {\bf 21}(12), \ 1155 \ (2019).   
\bibitem{ASW} S. Artstein-Avidan, S. Sadovsky, K. Wyczesany, \ \emph{A Zoo of Dualities}, \ J. Geom. Anal. {\bf 33}, \ 238 \ (2023).                                   
\end{thebibliography}
\end{document}